\documentstyle[prl,aps,epsfig]{revtex}
\begin{document}


 \twocolumn[\hsize\textwidth\columnwidth\hsize  
 \csname @twocolumnfalse\endcsname              

\title{Universal features of fluctuations}
 
\author
{Robert Botet$^{\dagger}$ and Marek P{\l}oszajczak$^{\ddagger}$}
 
\address{$^{\dagger}$
Laboratoire de Physique des Solides - CNRS, B\^{a}timent 510, Universit\'{e} Paris-Sud
\\ Centre d'Orsay, F-91405 Orsay, France  \\
and   \\  $^{\ddagger}$
Grand Acc\'{e}l\'{e}rateur National d'Ions Lourds (GANIL), \\
CEA/DSM -- CNRS/IN2P3, BP 5027,  F-14076 Caen Cedex, France  }
 
\date{\today}
 
\maketitle

\begin{abstract}
Universal scaling laws of fluctuations (the $\Delta$-scaling laws) 
can be derived for equilibrium and 
off-equilibrium systems when combined with the finite-size scaling analysis. 
In any system in which the second-order critical behavior can be identified, 
the relation between order parameter, criticality and scaling law of 
fluctuations has been established and the relation between the scaling 
function and the critical exponents has been found. 
\end{abstract}

\pacs{PACS numbers: 05.70.Jk,24.60.Ky,64.60.Ak,64.60.Fr,64.60.Ht}

 ]  

\section{INTRODUCTION}

Recent years have witnessed an intense experimental and theoretical activity
in search of scale-invariance and fractality in multihadron production
processes \cite{rep}. The creation of soft hadrons in these processes is
related to the strong-coupling, long-distance regime of Quantum Chromodynamics
(QCD) whose exploration remains the sacred grail of the high-energy particle
physics. Through a multitude of experimental studies of  particle density
fluctuations and particle correlations, new informations have been gathered
which prompted the suggestion to investigate the pattern of multiplicity
fluctuations in the small domains of phase-space looking for the
scale-invariance and fractality \cite{bialas}. 
New measurements has triggered revival of
interest in the theory of fluctuations in complex systems.

In this work we shall discuss universal features of fluctuations of physical quantities
in a $D$-dimensional, $N$-body system, where $N$ is essentially finite. 
We are particularly interested in the self-similar systems, 
such as the fractal objects or the thermodynamic systems at the second-order
phase transition. By the self-similarity we mean that 
the characteristic length $\sim (N^{*})^{1/D}$ 
($N^{*}$ is the characteristic size), which could be associated 
with the disappearance of fluctuations cannot be defined. 
The progress in this domain is associated with the  
development of the renormalization group methods. Actually,  
there is a close connection of the renormalization group ideas and the  
limit theorems in the probability theory. This basic idea stems from the  
works of Bleher, Sinai \cite{bleher73} and Jona-Lasinio \cite{jona75}, 
and consists of splitting the  
whole system into correlated blocks. Then the probability distribution of  
the investigated physical quantity is calculated for the blocks, and the  
renormalization procedure is expressing the distribution for large   
blocks in terms of the distributions for the smaller ones. These probability  
distributions dealing with larger and larger number of subunits   
are assumed to be given by probabilistic limit laws and the parameters of   
these distributions, properly renormalized at each step, are expected   
to tend to some finite limit. When this scheme is realized, 
the renormalization  
procedures gives the universal behavior of the infinite system.  

Following this approach, we shall discuss 
how the universal scaling laws of fluctuations of 
different observables appear in finite systems. In particular, 
we shall consider the order parameter
fluctuations in any equilibrium or non-equilibrium system in which the
second-order critical behavior can be identified. These considerations will 
provide an understanding
of the relation between the order parameter, the criticality and the 
scaling law of fluctuations. The notion of the relevant observable 
will also appear from this discussion.

\section{BROAD DISTRIBUTIONS AND LONG-RANGE CORRELATIONS}  
  
\subsection{A basic example of the probabilistic limit laws}  
  
Most physically interesting 'macroscopic variables' are defined as sum of   
many 'local random variables'. To understand such a  
'macroscopic behavior' is to seek for some universal limit behavior. 
The simplest, non-trivial example is  
when the local variables are uncorrelated. 
Suppose that we make $N$ trials and a certain  
event has finite probability $p_o$ to occur at each trial. In  
this chain of trials, we are looking for the   
behavior of the probability distribution $P_N[m]$  
to get exactly $m$ events during $N$ trials. 
In the following, $m$ will be called the multiplicity.  
This kind of process could describe for example 
the successive emission of particles : at each time interval $\tau$, 
one particle is emitted with a probability $p_o$ from a parent body. 
The equation for the probability distribution of the number of 
emitted particles in the time interval $n\tau$ :  
\begin{eqnarray}
\label{nonM}  
P_N[m] = (1-p_o) P_{N-1}[m] + p_o P_{N-1}[m-1]   \ ,  
\end{eqnarray}  
tells that the multiplicity at stage $N$ is equal to $m$ if, either 
it was $m$ at stage $N-1$ and then the emission does not 
occur at the following trial,  
or it was $m-1$ at stage $N-1$ and the emission occurs in the last trial. 
In particular, eq. (\ref{nonM}) shows that the average multiplicity is   
$<m> = Np_o$. When the number of trials $N$ becomes large, 
the  probability distribution $P_N[m]$ tends to a normal law :
\begin{eqnarray}
\label{2scaling}  
& &\lim_{<m> \rightarrow \infty} <m>^{1/2} P_N[m] {\longrightarrow}   
\nonumber \\ &\longrightarrow&
a_{p_o} \exp \left[ -b_{p_o} 
\left(\frac{m-<m>}{<m>^{1/2}}\right)^2 \right]  
\  ,  
\end{eqnarray}  
with $b_{p_o}=1/[2(1-p_o)]$. This form   
of the normal law expresses a relation between the scaled distribution 
$<m>^{1/2} P_N[m]$, and the shifted scaled multiplicity parameter
$(m-<m>)/<m>^{1/2}$.

Let us now introduce the correlation between the two consecutive 
trials by prohibiting the successive emission. 
In this case, the probability distribution verifies :  
\begin{eqnarray}
\label{Mark}  
P_N[m] = (1-p) P_{N-1}[m] + p P_{N-2}[m-1]   \ .
\end{eqnarray}  
It is a simple exercise to obtain the solution of  
eq. (\ref{Mark}) : the leading order of the average 
multiplicity is $<m> \sim {np}/(1+p)$ \cite{comment} and
the limit distribution is given again by a normal law (\ref{2scaling}) 
with $b_{p_o}=1/[2(1-p_o)(1-2p_o)]$. Hence, the 
short-range correlations in (\ref{Mark})
reduce the variance of the probability distribution by a factor $1-2p_o$.
It is important to realize a fundamental difference between the limit law, 
which is universal in the sense that most details of the considered 
process are unessential, and the parameters of the limit   
laws, which strongly depend on the specific features of the process.

\subsection{Central limit and stable laws for non-correlated variables}  
  
Let us consider the additive, 'global' quantity :
$Y_N = \sum_{n=1}^{N} X_n$,
expressed in terms of 'local' microscopic variables $X_n$.  
The fundamental problem in the statistical description of composed objects 
is to find a  limit distribution of this variable, 
with $N$ being the number of subunits of the system. This problem can be 
approached in two steps : (i) when the  
random variables $X_n$ are all independent, then 
(ii) when they are correlated.  
The first step has been  solved entirely \cite{levi}. The second step, when  
the variables are correlated, is not yet completely solved but some precise  
results and tools are available (see Ref. \cite{math} for recent references).  
  
In the following, we will work with the probability density :
$f(x)dx=\mbox{Prob}[x \le X < x+dx]$, its characteristic function : 
\begin{eqnarray}
\phi_X(k) = \int_{-\infty}^{\infty} e^{ikx} f(x) dx \nonumber  \ ,
\end{eqnarray} 
and the expansion of the characteristic function 
in a series of cumulants near $k=0$ :   
\begin{eqnarray}
\label{lnphi}  
\ln \phi_X(k) & = & ik<X>-  \nonumber \\ &-&\frac{k^2}{2}(<X^2>-<X>^2)+ \dots  
\end{eqnarray}  
The importance  of the characteristic function comes 
from the convolution theorem which tells that, if $X_1$ and $X_2$ are  
two independent random variables and $\phi_{X_1}(k)$ and $\phi_{X_2}(k)$ their 
characteristic functions,  
then the characteristic function of the random variable $Y=c_1X_1+c_2X_2$ is :  
$\phi_Y(k)=\phi_{X_1}(c_1k) \phi_{X_2}(c_2k)$ 
($c_1$ and $c_2$ are real numbers). In other words,
the density of the sum of two independent variables is the product of the 
convolution of the two probability densities.  
The key point to determine all possible 
limit distribution of the random additive variable is the 'stability problem' : 
{\it What are all probability densities $f(x)$,   
such that, if $X_1$ and $X_2$ are two independent 
random variables of a probability  density $f(x)$, and $c_1$  
and $c_2$ are two real numbers, then $Y=c_1X_1+c_2X_2$ is a random variable  
with {\it the same} probability density $f(x)$ ?} 
It turns out that the stable distributions are 
those for which the logarithm of their   
characteristic function is given by the formula : 
\begin{equation}
\ln \phi(k)=i \gamma k-c |k|^{\mu} \left(1+i \beta \frac{k}{|k|} 
\omega_{\mu}(k)   \right) ~ \ ,
\end{equation} 
with four real parameters : $\gamma , c, \beta $ and $\mu$, where
$c > 0$, $ |\beta | \le 1$ and $0 < \mu \le 2$. In addition : 
$ \omega_{\mu}(k) = \tan(\pi \mu/2)$  
if $\mu \neq 1$, else $\omega_{1}(k)=2\ln (k /\pi)$. This four-parameter 
family of functions can be written in different forms.
One of them is :  
\begin{eqnarray}  
\label{Levy1}  
&&P_{\mu ,\beta }(c^{1/\mu }x+\gamma ) = \frac{1}{\pi c^{1/\mu }} \times
\nonumber \\&\times &{\mbox{Re}}
\left[ \int_{0}^{\infty }  
\exp \left( -ikx -(1+i\beta \tan \frac{\pi \mu }{2})k^{\mu }\right) dk \right] 
\nonumber
\end{eqnarray}
if $\mu \neq 1$, and
\begin{eqnarray} 
&&P_{1,\beta }(cx+\gamma ) = \frac{1}{\pi c} \times \nonumber \\ 
&\times&\int_{0}^{\infty }  
\exp (-k)\cos \left( k(x+\frac{2\beta }{\pi }\ln k)\right) dk   \nonumber
\end{eqnarray}
if $\mu = 1$. For $|x| \rightarrow \infty$,
$P_{\mu,\beta} \propto |x|^{-(1+\mu)}$ ~($0 < \mu < 2$). For $0 < \mu < 1$
and $\beta \neq 0$, there is an essential singularity when $x \rightarrow 0$ :
$P_{\mu,\beta} \propto \exp [-\alpha(\mu,\beta)x^{\mu/(1-\mu)}]$. For
other values, $P_{\mu,\beta}$ is finite at the origin. 

The parameters appearing in these formulas are not equally important. 
$\gamma$  corresponds to a shift of the whole distribution, 
while $c$ is a scale factor. These two parameters are 
related to a translation and a dilatation  
of the distribution and can always be put respectively equal to 0 and 1 by a   
suitable choice of the coordinates. 
The parameter $\beta$ is related to the asymmetry of the distribution
and $\beta$ vanishes for an even distribution.  
The most important parameter is the characteristic exponent   
$\mu$. It cannot be larger than 2 because 
$P_{\mu,\beta}$ should be non-negative, and is   
larger than 0 to ensure the convergence at the origin. 
The normal (gaussian) distribution corresponds  
to $\mu=2$, and this is the only 
case where the variance $\sigma_X \equiv <X^2>-<X>^2$ is finite. In fact,  
using (\ref{lnphi}) and the form
of the stable distributions, one may show that :  
\begin{eqnarray}  
0 < \mu \leq 1 & \longrightarrow & <X>=\infty ~~{\mbox {and}}~~ 
\sigma_X=\infty \nonumber \\  
1 < \mu < 2 & \longrightarrow & <X><\infty ~~{\mbox {and}}~~ 
\sigma_X=\infty \nonumber \\  
\mu=2 & \longrightarrow & <X><\infty ~~{\mbox {and}}~~ 
\sigma_X<\infty \nonumber \ .  
\end{eqnarray}  
  
\subsection{Asymptotically stable laws  - the domains of partial attraction}  
  
Let us recall that the stable distribution is invariant when added to itself.   
One would like to know what kind of distributions of the local random variable $X_n$
are leading to the  
stable distribution $P_{\mu,\beta}$ when $N \rightarrow \infty$.  
Let $X_1, X_2, ... , X_N$ denotes a sequence  
of independent numbers distributed according to a  
density $f(x)$, and let  $Y_N$ be a random variable : 
\begin{eqnarray}
\label{defvar1}
Y_N = X_1+\dots+X_N \hspace{0.7cm} 
\mbox{if} \hspace{0.3cm} 0 < \mu < 1 \hspace{0.3cm} ~ \ ,   
\end{eqnarray} 
and
\begin{eqnarray}
\label{defvar2}
Y_N = X_1+\dots+X_N-N<X> \hspace{0.2cm} ~ \ . 
\end{eqnarray} 
if $1 < \mu < 2$. Let us define : ${\cal Y}_N=Y_N/B_N$,
with the parameter $B_N$ chosen in such a way
that the probability distribution of $\cal{Y}_N$ converges to the   
stable distribution $P_{\mu,\beta}$ when $N$ tends to the infinity. When the   
distribution $f(x)$ yields the limit density $P_{\mu,\beta}$  
for $\cal{Y}$, then $f$ is said to belong to the domain of attraction 
of $P_{\mu,\beta}$.  
The ensemble of the probability densities $f(x)$, such that :  
\begin{eqnarray}
\label{bassin}  
\lim_{x \rightarrow \infty} & &
\frac{\int_x^{\infty} f(x) dx+\int_{-\infty}^{-x} f(x') dx'}  
{\int_{ax}^{\infty} f(x) dx+\int_{-\infty}^{-ax} f(x') dx'} 
\sim a^{\mu} \nonumber \\ \nonumber \\  \nonumber \\
\lim_{x \rightarrow \infty} & &
\frac{\int_{-\infty}^{-x} f(x') dx'}{\int_x^{\infty} f(x')   
dx'}= \frac{1-\beta}{1+\beta} \nonumber ~ \ ,  
\end{eqnarray}  
constitutes the entire domain of attraction of $P_{\mu,\beta}$. 
We see that only the tail of the distribution $f(x)$ is significant.
If $f(x) \sim |x|^{-(1+\mu)}$ ~~($0<\mu<2$)  
for $|x| \rightarrow \infty$, then $f(x)$ belongs 
to the domain of attraction of  $P_{\mu,\beta}$, {\it i.e.}
the reduced global variable $\cal{Y}_N$ has a limiting stable   
distribution of index $\mu$ when $N \rightarrow \infty$.  
Moreover, all parameters $B_N, \gamma, c, \beta$ are known explicitly.  
In particular, $B_N \sim N^{1/\mu}$.

\subsection{Stable distributions and the self-similarity}

The stochastic process $Y_N$ is called {\it self-similar} 
with the Hurst exponent $H$ if both
$Y_{\lambda N}$ and ${\lambda}^HY_N$ have the same distribution. 
If $Y_N$ is given by (\ref{defvar1}), (\ref{defvar2}), respectively, and the
probability distribution $P[\cal{Y}_N]$ converges to
the stable law $P_{\mu,\beta}$, then the stochastic process $Y_N$ is
self-similar with the Hurst exponent $H=1/\mu$, and 
$<Y_N> \sim N^{1/\mu} = N^g$. $g$ is the anomalous dimension related  to the
extensive variable $Y_N$.
  
\subsection{Concept of the $\Delta$-scaling} 
  
The case : $1<\mu<2$, corresponds to an infinite  
variance of microscopic variable $X$ with finite  mean value $<X>$. 
Let $M_N=X_1+...+X_N$. In this case :
\begin{eqnarray}
\label{glevi}
<M_N> \sim <X_1>+...+<X_N> = N<X> \nonumber ~ \ .
\end{eqnarray}  
The preceding characterization of the 
attraction domains of the stable laws yields : 
\begin{eqnarray}  
\mbox{Prob}\left[ y \leq \frac{M_N-N<X>}{B_N} <y+dy \right] \sim P_{\mu,\beta}(y)dy \nonumber 
\end{eqnarray}
 with   $B_N \sim <M_N>^{1/\mu}$. 
Introducing the probability density $f_{M_N}(m)$ of the  
variable $M_N$, one obtains the limit distribution :  
\begin{eqnarray}
\label{deltaMN}  
<M_N>^{1/\mu} f_{M_N}(m)\sim
P_{\mu,\beta}\left(\frac{m-<M_N>}{<M_N>^{1/\mu}}\right)  
\end{eqnarray}  
, when the number $N$ of subunits $X_n$ tends to 
the infinity. Notice that $P_{\mu,\beta}$ appears now as the 
{\it scaling function} with the scaling index $\Delta = 1/\mu$, 
which can be extracted  by plotting $<M_N>^{1/\mu}f_{M_N}(m)$
against a scaling variable $(m-<M_N>)/<M_N>^{1/\mu}$. In this representation,
different curves for systems or stochastic processes 
with different number of subunits
$N$ collapse into a single scaling curve which characterizes the stable law. 
The scaling law (\ref{deltaMN}) 
will be called in the following the {\it $\Delta$-scaling law}. 
We shall return to this form later
when discussing the second-order critical behavior in finite systems. 
  
The case $0<\mu<1$ is more subtle, because $<M_N>$ cannot be defined. 
In this case, a limiting behavior is expected to take a form :   
\begin{eqnarray}
\label{yyy}
N^{1/\mu} f_{M_N}(m) \sim P_{\mu,\beta}\left(\frac{m}{N^{1/\mu}}\right) ~ \ .
\end{eqnarray}
There is no connection between $<M_N>$ and $N$ in (\ref{yyy}). 
Instead, one can show that the typical value of $M_N$ for large values of $N$ is
: $M_N^{*} \simeq N^{1/\mu}$, so the correct scaling is :
\begin{eqnarray}
\label{xxx}
M_N^{*} f_{M_N}(m) \sim P_{\mu,\beta}\left(\frac{m}{M_N^{*}}\right) \ .
\end{eqnarray}  
Notice that (\ref{xxx}) resembles the limit when $\mu \rightarrow 1$ of the 
scaling law (\ref{deltaMN}). 
This scaling form is not very useful because $M_N^{*}$ is not 
as precisely defined as the average value.   
It shows however, that it does not make sense to search for the scaling relations such
as (\ref{deltaMN}) with the scaling exponent $1/\mu$   larger than 1.  
  
\section{ORDER PARAMETER FLUCTUATIONS} 

Fluctuations of the order parameter 
are expected to have different properties at the critical point 
and outside of it. Far from the critical point, the correlations
are short-ranged and the results of the previous chapter apply in
this case. On the contrary, 
fluctuations of the order parameter close to the critical point are correlated 
throughout the whole system. Hence, the 'macroscopic'
order parameter cannot be written as a simple sum of uncorrelated 
block variables, and its
asymptotically stable distribution
differs from those discussed in chapt. 2. In the next
section we shall discuss some features of these distributions and show that
they satisfy the $\Delta$-scaling law.

\subsection{Finite-size scaling argument in the 
thermodynamical systems}

In the thermodynamic limit, the free energy density of an equilibrium system 
close to the critical point scales as \cite{Widom} :
\begin{eqnarray}
f (\lambda^{\beta} \eta, \lambda \epsilon) \sim 
\lambda^{2-\alpha} f (\eta, \epsilon) ~ \ ,
\end{eqnarray}
where $\alpha, \beta $ are the usual critical exponents, $\eta$ is 
the intensive order parameter and $\lambda$ is the scale parameter. $\epsilon$
is the control parameter which equals 0 at the critical point.
Even though finite systems do not exhibit the critical behavior, 
nevertheless their properties resemble those of infinite systems 
if the correlation length $\xi $ is larger or comparable to the typical 
length $L$~ of the system. In this case, one speaks about the pseudocritical 
point at a distance : 
\begin{eqnarray}
\label{row1}
\epsilon \sim c N^{-1/{\nu D}} \nonumber
\end{eqnarray}
from the true critical point \cite{FF}. $N$ is the size of the
$D$-dimensional system and $c$~ is some dimensionless constant which
is negative, if a maximum of the finite-size 
susceptibility lies in the ordered phase, or positive, 
if this maximum is in the disordered 
phase. One can then derive the finite-size scaling of the total free energy 
$F(\eta , \epsilon, N) = N f (\eta , \epsilon )$ at the pseudocritical point :
\begin{eqnarray}
\label{Widomfinite}
F_c(\eta ,N) \sim f(\eta N^{\frac{\beta }{2-\alpha}}) ~ \ .
\end{eqnarray}
In deriving (\ref{Widomfinite}), we used the hyperscaling relation : 
$2-\alpha =  \nu D ~ \ . $ The canonical probability density of the order 
parameter $P_N[\eta ]$ is given by :
\begin{eqnarray}
\label{psi}
P_N[\eta ] = {Z_N}^{-1} \exp[-{\beta}_T F(\eta , \epsilon , N)] ~ \ ,
\end{eqnarray}
where the coefficient ${\beta}_T ( \equiv 1/T )$ is independent of $\eta $ 
($T$ is the temperature of the system). Using eq. (\ref{psi}), 
one may calculate not only the most probable value of the order parameter, 
which is the solution to the equation  
$\partial P_N[\eta]/\partial \eta = 0$,
but also the average value of the order parameter and the partition function 
\begin{eqnarray}
\label{part}
Z_N \sim N^{-\frac{\beta }{2-\alpha}} \sim ~<\eta >~ \sim ~\eta^{*} ~ \ .
\end{eqnarray}
${\eta}^{*}$ denotes the most probable value of the order
parameter. $<\eta>$ vanishes for large values of
$N$, since both $\beta$ and $2-\alpha = 2 \beta + \gamma$ are positive.
The probability density $P_N[\eta ]$ obeys {\it the first-scaling law} :
\begin{eqnarray}
\label{first}
<\eta > P_N[\eta ] = \Phi \left( \eta/< \eta > \right) \equiv \Phi (z) 
\end{eqnarray}
with
\begin{eqnarray}
\label{firstsc}
\Phi (z) \sim \exp \left( - {\beta}_T f \left(a z, c \right) \right)  ~ \ .
\end{eqnarray}
The logarithm of scaling function $\Phi (z)$ corresponds to the 
non-critical free energy density at the renormalized distance $\epsilon =c$~ 
from the critical point. The scaling law
(\ref{first}), which is analogous to (\ref{xxx}), can be rewritten as :
\begin{eqnarray}
\label{stand1}
<m> P_N[m] = \Phi(z_{(1)}) ~ \ ,
\end{eqnarray}    
where $m = N \eta$ is the extensive order parameter and $z_{(1)}$ is :
\begin{eqnarray}
\label{var1}
z_{(1)} = (m-m^{*})/<m> ~ \ .
\end{eqnarray}
$m^{*}$ denotes the most probable value of $m$.
The scaling domain is defined by the asymptotic behavior of $P_N[m]$
when $m \rightarrow \infty$ and $<m> \rightarrow \infty$, but $z_{(1)}$ 
has a finite value. Assuming that the scaling framework of the 
second-order phase transition 
holds, the scaling relation (\ref{first}) is valid independently of the
explicit reasons of changing $<m>$, and independently of any phenomenological 
details. In other words, an explicit relation between  the size $N$ of the
system and $<m>$ need not to be known. 

If the order parameter is related to the multiplicity of 
fragments or produced particles, like in the 
fragmentation - inactivation - binary (FIB) 
process \cite{sing11}, then the first scaling (\ref{first}) becomes identical 
to the well known KNO-scaling \cite{koba}. Only in this case, 
the multiplicity of produced particles is the relevant observable.

\subsection{Tail of the scaling function}

What are the properties of the scaling function for large values of $m/<m>$? 
The correct way of approaching this problem is
to study the system subject to a small field $h$ which is conjugate to the order
parameter and breaks the symmetry of the distribution. 
This consideration yields \cite{botnowa} :
\begin{eqnarray}
\label{tail}
\Phi(z) \sim \exp ({-a z_{(1)}^{\delta + 1}}) \equiv \exp ({-a z_{(1)}^{\hat \nu}}) ~ \ ,
\end{eqnarray}
with $\delta = (2-\alpha-\beta)/{\beta}$. The coefficient $a$ in (\ref{tail}) 
depends regularly on the temperature. 

One can express this relation in a different way. The anomalous dimension for  
an extensive quantity $m = N \eta$ can be defined as : 
\begin{eqnarray}
\label{anomdim}
g = \lim_{N \rightarrow \infty} g_N = 
\lim_{N \rightarrow \infty} \frac{d}{d\ln N} \left( \ln <m> \right) ~ \ . 
\end{eqnarray}
One can see from (\ref{part}) and the Rushbrooke relation : 
$\alpha+2 \beta+ \gamma = 2 ~ \ ,$ that the anomalous dimension is :  
\begin{eqnarray}
\label{an1}
g=1- \beta/(\gamma + 2 \beta) ~ \ . 
\end{eqnarray}
Since $\alpha$ and $\beta$ are both positive, therefore 
$g$ is contained between 1/2 and 1 for equilibrium 
systems at the critical point of 
the second-order phase transition. One may note also that :
\begin{eqnarray}
\label{gdab}
{\hat \nu} = \frac{1}{1-g} = \frac{2-\alpha}{\beta} > 2 ~ \ .
\end{eqnarray}

Whenever the cluster-size can be correctly defined 
, like for example in the case of percolation or Ising models,
the exponent $\tau$ of the power-law cluster-size distribution  
$n_s \sim s^{-\tau }$ satisfies : 
\begin{eqnarray}\
\label{g}
g = \frac{1}{\tau -1} \hspace{0.5cm} \mbox{and} \hspace{0.5cm}
{\hat \nu} = \frac{\tau -1}{\tau -2} ~ \ .
\end{eqnarray}
The allowed values at the critical point are : $2 < \tau < 3$.
Consequently, one can learn from the cluster-size distribution 
whether the equilibrium system is at the critical point
and whether the considered extensive quantity can be identified with 
the order parameter. For this kind of equilibrium phase transitions 
, the size of the largest cluster is the order parameter 
since : $$<s_{max}> \sim N^{\frac{1}{\tau-1}} \equiv N^g\ .$$
The cluster multiplicity
could be the order parameter if $\tau < 2$, but this can be satisfied only
for certain off-equilibrium phase transitions, such as for example the
shattering phase transition in the FIB process 
\cite{sing11} or in the dissipative perturbative QCD \cite{zphys}. 

\subsection{The $\Delta$-scaling law}

What happens if the observable $m$ is the $N$-dependent 
function of the order parameter : 
\begin{eqnarray}
\label{delta0}  
m=N^{a_1 } - \eta ~~ \mbox{with}~~ a_1 > g ~ \ . 
\end{eqnarray}
Writing (\ref{stand1}) with 
$m$ instead of $\eta$ and taking into account that 
$P_N[\eta ] d\eta =P_N[m]dm$,  one finds :
\begin{eqnarray}
\label{delta}
<m>^{\Delta}P_N[m] = \Phi (z_{({\Delta})}) ~ \ ,
\end{eqnarray}
with the scaling variable :
\begin{eqnarray}
\label{deltawar}
z_{(\Delta )} = (m-m^{*})/<m>^{\Delta} ~ \ ,
\end{eqnarray} 
where : 
\begin{eqnarray}
\label{deltarel}
\Delta=g/a_1 < 1  ~ \ . 
\end{eqnarray}
The normalization of $P_N[m]$ and the definition of $<m>$ provide the 
two constraints which are consistent with : $0 < \Delta \le 1$. 
This $\Delta$-scaling law ($0 < \Delta < 1$) 
is satisfied, as shown in sect. 2.5, by the
asymptotically stable distributions with $1 < \mu < 2$ (eq. (\ref{deltaMN})).
The asymptotically stable distributions $P_{\mu,\beta}$ 
with $0 < \mu < 1$ satisfy the first
scaling law, {\it i.e.} the limiting case of the $\Delta$-scaling for $\Delta =
1$. 

The scaling function ${\Phi}(z_{(\Delta )})$ has the
identical form as ${\Phi}(z_{(1)})$, except for the inversion of the 
abscissa axis. In particular, its tail for large $z_{(\Delta )}$ has the same form :
\begin{eqnarray}
\label{deltatail}
\Phi (z_{(\Delta )}) \sim \exp \left( -z_{(\Delta )}^{{\hat \nu}} \right) = 
\exp \left( -z_{(\Delta )}^{\frac{1}{1-g}} \right)  
\end{eqnarray}
as given in (\ref{tail}). Notice that the 
$\Delta $-scaling of an extensive variable : 
${\hat m} = N(1 - \eta ) \equiv N{\hat {\eta}}$,
can be used to determine directly the anomalous 
dimension, since in this case : $\Delta = g$. 
At the phase transition : $< N{\hat {\eta}} >~ \sim N$, but the 
finite-size corrections are {\it algebraic}. 
One should mention in passing that if 
$a_1 < g$, then $<m> \sim N<\eta >$ and $\Delta = 1$. 

It is important to realize that the $\Delta$-scaling ($0<\Delta < 1$) and the
first-scaling ($\Delta = 1$) laws are
satisfied by the distributions of the quantities which are the sum of either
independent or correlated random variables. In this sense, these 
scaling laws are the fundamental characteristics of the composed statistical
system. 

\subsection{Cumulant relations}  

Let us now introduce the cumulants $\kappa_q$ from the formula 
for the moment expansion of the probability distribution $P[m]$ :
\begin{eqnarray}
\label{cum}
\ln \left( \sum_{m=0}^{\infty} P[m] \exp (m u) \right) = \sum_{q=0}^{\infty}
\frac{u^q}{q!} \kappa_q ~ \ .
\end{eqnarray}
In the case of $\Delta$-scaling law (eq. (\ref{delta})), 
the cumulants satisfy : 
\begin{eqnarray}
\label{cumulrel}
\kappa_q \sim <m>^{q\Delta} \equiv <m>^{qH} \hspace{0.5cm} (q \geq 2)~ \ .
\end{eqnarray} 
In the case of the first-scaling law (eq. (\ref{stand1})), one finds : 
\begin{eqnarray}
\label{cumulrel1}
\kappa_q \sim <m>^q ~ \ . 
\end{eqnarray}
Notice also that the scaled variance $\kappa_2/\kappa_1^2$ 
is independent of $N$ for $\Delta = 1$ and tends to 0 
when $N \rightarrow \infty$ for $0 < \Delta < 1$. In
this sense, the scaling law (\ref{stand1}) is the large fluctuation limit of
the probability distribution $P_N[m]$.

\subsection{The intermittent behavior}

Intermittency in particle physics is defined usually as the scale-invariance of
the factorial moments with respect to the change of the size of the phase-space
cell $\delta y$ :
\begin{eqnarray}
\label{fact}
F_q(\delta y) \propto (\delta y)^{-\phi_q} ~ \ ,
\end{eqnarray}
when $\delta y \rightarrow 0$. The exponent $\phi_q$ is called the
intermittency index. For the multifractal distribution we have :
\begin{eqnarray}
\label{multifr}
\sum_ip_i^q(\delta y) = <p_i^{q-1}(\delta y)> \propto (\delta y)^{\tau(q)} ~ \ ,
\end{eqnarray}
where $p_i(\delta y)$ is the probability for the particle 
to be in the bin $i$ of size $\delta y$ and $\tau(q)$ is given by :
\begin{eqnarray}
\label{tau}
\tau(q) = (q-1)D_q\Delta ~ \ .
\end{eqnarray}
Since :
\begin{eqnarray}
\label{momfac}
F_q \sim (\delta y)^{-(q-1)D}<p_i^{q-1}> ~ \ ,
\end{eqnarray}
where $D$ is the support dimension, therefore :
\begin{eqnarray}
\label{phix}
\phi_q = (q-1)(D-D_q\Delta) ~ \ .
\end{eqnarray}
The quantity $D_q\Delta$ appears 
here as the apparent Renyi dimensions. Hence, knowing the
intermittency indices and the value of $\Delta$, 
one can extract the 'true' multifractal spectrum.

\subsection{Scaling laws and the self-similarity}

The self-similarity of the statistical system which obeys the first-scaling law
(eqs. (\ref{stand1}), (\ref{var1}))
means that : $$<N\eta>_{\lambda N} \sim \lambda^g<N\eta>_N \equiv
\lambda^H<N\eta>_N ~ \ .$$ The
Hurst exponent is equal to the anomalous dimension $g$ (eq. (\ref{anomdim})). 
In the case of the
$\Delta$-scaling, $H\equiv g=a_1\Delta$ (see eq. (\ref{deltarel})). 
The exponent $a_1$ is the apparent anomalous
dimension for $m$ defined by (\ref{delta0}) : $<m>_{\lambda N}\sim {\lambda}^{a_1}<m>_N$. 
It is easy to see, that these properties are
analogous to those we have discussed in sect. 2.4 with $Y_N$ corresponding to
$N\eta$ and $\sum X_i$ corresponding to $m$. Namely : 
$$a_1=1,~~ g=1/\mu,~~ \Delta = 1/\mu$$ for $1 < \mu < 2$, whereas 
$$a_1\equiv g=1/\mu,~~ \Delta = 1$$ for $0 < \mu < 1$.

\subsection{The off-critical scaling}

Away from the critical region, if $m$ is the order parameter then 
the finite system exhibits 
{\it the second-scaling law} (the limit $\Delta = 1/2$ of the $\Delta$-scaling
law (\ref{delta}), (\ref{deltawar})) in the ordered phase, and the first-scaling law 
(eqs. (\ref{stand1}, (\ref{var1})) 
in the disordered phase. In both cases, the tail of the scaling 
function is gaussian (${\hat \nu}=2$) and the finite-size corrections are 
{\it exponential}. The second-scaling law has been found in the shattering 
phase of the non-equilibrium FIB process \cite{nowa} and 
in the 'liquid' phase of the equilibrium percolation process \cite{prln}. The
first-scaling law with the gaussian tail has been found in the disordered
phase of the Ising model \cite{bouchaud}.
 
One would like to know how a finite 
system evolves close to the critical point when the control parameter 
$\epsilon$ tends slowly to 0 : 
$\epsilon \sim N^{2a_1 -2} ~~(\epsilon < 1)$ \cite{botnowa}. In this case, 
$<m>$ and $m^{*}$ have a common behavior $\sim N^{a_1}$, and the probability
distribution $P_N[m]$ satisfies the scaling :
\begin{eqnarray}
\label{devia}
<m>^{\Delta} P_N[m] \sim \exp \left( -c \frac{(m-m^{*})^2}{<m>^{2\Delta}}
\right) 
\end{eqnarray}
with $c > 0$ and $\Delta = 3a_1/2 - 1  \ .$
The case $a_1 = 1$, {\it i.e.}, $\epsilon = {\mbox {const}}$, 
corresponds to the second-scaling law. 
$a_1 = 3/4$ corresponds to the first-scaling 
law since the finite system is in 
the critical region ($a_1 = g$). In between these two limiting cases, 
the $\Delta$-scaling 
holds with $1/2 < \Delta < 1$. In all cases, the
scaling function has a gaussian form.

\subsection{Equilibrium vs off-equilibrium critical systems}

Let us consider the bond percolation model. 
In the regular lattice, each site corresponds to a monomer and 
a proportion $p$ of active bonds  between occupied sites is set randomly.
Such a network results in a distribution of clusters which are defined as 
the ensemble of occupied sites connected by active bonds. 
For a critical value of active bonds 
$p_{cr}$, an 'infinite' cluster (the gel) spans the whole 
lattice. Infinite in this context means that the gel contains 
a finite fraction of the total mass of the system. The sol - gel phase 
transition can be suitably studied using moments of the 
number-size-distribution $n_s$ :
$M_k' = \sum_{s < s_{max}} s^k n_s  ~ \ ,$
where the sum runs over all clusters with the exception of the 
largest cluster $s \equiv s_{max}$ and the superscript $'$~ recalls this 
constraint. The mass of the largest cluster 
is then : $N-M_1'$. In the infinite system, one usually works with the 
normalized moments of the concentration-size-distribution $c_s$ :
$m_k' = \sum_{s<s_{max}} s^k c_s  ~ \ .$
The concentrations are normalized  such that :
$c_s= \lim_{N \rightarrow \infty} n_s/N ~$.
The probability that a monomer belongs to the infinite cluster 
(the gel) is equal to $1-m_1'$ with : 
$m_k'= \lim_{N \rightarrow \infty} M_k'/N ~$.
Therefore, in the thermodynamic limit : $m_1' = 1$ for $p < p_{cr}$ and 
$m_1' < 1$ for $p > p_{cr}$ , {\it i.e.}, a finite 
fraction of the total number of sites belongs to the infinite cluster.

After this introduction, let us discuss the bond-percolation 
on the Bethe lattice with a  coordination number ${\hat z}$ \cite{Fisher}. 
The concentration-size-distribution in this model is :
\begin{eqnarray}
\label{eq4}
c_s ={\hat z}\frac{(({\hat z}-1)s)!}{(({\hat z}-2)s+2)! s!} p^{s-1} 
(1-p)^{({\hat z}-2)s+{\hat z}}  
\end{eqnarray}
and the first normalized moment equals :
\begin{eqnarray}
\label{eq5}
m_1'= \left( \frac {1-p}{1-p^{*}} \right)^{2{\hat z}-2} ~ \ ,  \nonumber 
\end{eqnarray}                          
with $p^{*}$ being the smallest solution of the equation : 
\begin{eqnarray}
\label{eq6}
p^{*}(1-p^{*})^{{\hat z}-2} =p(1-p)^{{\hat z}-2}  ~ \ .
\end{eqnarray}
For $p < p_{cr} = ({\hat z}-1)^{-1}$, the only solution of eq. (\ref{eq6}) is : $p^{*} = p$, 
but when $p$ is larger than $p_{cr}$, then there exists
a smaller non-trivial solution which behaves as 
$p_{cr}-|p-p_{cr}|$ near $p_{cr}$. Above 
this threshold, the moment $m_1'$ is smaller than 1 
and behaves approximately as : 
$m_1^{'} \simeq 1-2(p-p_{cr})/(1-p_{cr})~ \ .$ 

For large values of the size $s$, the concentrations 
(\ref{eq4}) can be rewritten using the 
Stirling approximation :
\begin{eqnarray}
\label{eq6a}
c_s \sim s^{-5/2} \exp (-\alpha s) ~ \ . 
\end{eqnarray}
$\alpha $ in (\ref{eq6a}) equals :
\begin{eqnarray}
\label{eq8}
\alpha &=& \ln \left[ \frac {p}{p_{cr}} \left( \frac {1-p}{1-p_{cr}} 
\right)^{{\hat z}-2}  \right] \nonumber \\
&\simeq& ({\hat z}-2)\ln (1-p) < 0 ~  \ .
\end{eqnarray}
Hence : $$c_s \sim s^{-5/2} (1-p)^{-s({\hat z}-2)} ~ \ .$$ 
At the percolation threshold, the concentrations follow the power-law and
outside this threshold, an exponential cut-off is always present. 

If we impose that $s_{max}<\infty$, even in the thermodynamical limit, 
then the activation $p$ corresponding to $s_{max}$ is smaller than $p_{cr}$ 
and the limiting activation $p_0$ equals
$p_{cr}$. The limiting activation $p_0$ in percolation is analogous to the 
limiting temperature $T_0$ in the statistical bootstrap model (SBM) 
\cite{bootstrap}. Both $p_0$ and $T_0$ follow from the  
consistency conditions in the two models and describe the asymptotic 
mass spectra. Obviously,
the limiting value $p_0$/$T_0$ of
the control parameter $p$/$T$ does not have to correspond to the critical
point in the equilibrium system. 
For this to be true, the exponent $\tau$ of the mass spectrum must be contained
in between 2 and 3.

One needs two critical exponents to describe completely the critical features. 
In percolation these are
for example $\tau $ and $\sigma $ (the exponent of the mean cluster-size 
divergence). The
$\Delta$-scaling analysis allows to determine only this exponent which is 
related to the anomalous dimension (e.g. $\tau$ in the percolation problem).

 The insights gained from the numerical simulations of non-reversible
aggregation equations (the Smoluchowski model) and FIB
equations, both describing the off-equilibrium phenomena, 
and the evidences from analytical solutions for gelling and 
non-gelling systems \cite{botnowa}, provide strong indication that the
scaling features, which were discussed in chapt. 2 for the distribution of 
the sum of random short-ranged correlated variables
and in chapt. 3 for the distribution of the 
long-ranged correlated variables at critical point in equilibrium systems, 
are valid also for non-equilibrium systems. This
universal behavior of equilibrium and non-equilibrium systems, 
has the deep foundation in the relation between renormalization group ideas for
self-similar systems and the limit theorems of the probability theory 
for the $\Delta$-scaling laws, both for correlated and
uncorrelated variables. The concept
of statistical equilibrium does not intervene at this level. 
\noindent
\section{WHERE WE ARE, WHERE DO WE GO?}

There are two generic classes of dynamical critical phenomena, which are 
characterized
by different 'relevant observables'. The first one is the sequential cluster
fragmentation, where the average cluster size decreases 
and the cluster multiplicity increases during the process. The cluster 
(particle) multiplicity is an order parameter for this class of critical
phenomena. The second one is the sequential cluster aggregation, where the
average cluster size increases and the cluster multiplicity decreases
during the process. The size of the largest cluster is here an order parameter.
In the first class, one finds for example 
the shattering phase transition in the FIB process \cite{sing11} 
as well as in the dissipative perturbative QCD process
\cite{zphys}. The second class contains for example the liquid-gas  
phase transition (Fisher model), the percolation transition or the sol-gel
phase transition in the irreversible kinetic aggregation. Which of them is
approximately realized in the multiparticle production process, 
can be discovered by
studying the $\Delta$-scaling law, the tail of the scaling function, and the
anomalous dimension, as described in the previous chapters.
 
\subsection{The perturbative QCD equations with the inactivation mechanism}

Let us call $P_N[n;t]$ the probability
to get a cluster multiplicity $n$ at time $t$,
starting from an initial cluster of size $N$ at $t = 0$. The time
evolution equation for the multiplicity is given by the 
non-linear rate equation \cite{sing11} :
\begin{eqnarray}
\label{no1}
\frac{\partial G_N}{\partial t}(u,t)&=&
\sum_{j=1}^{N-1} {\cal F}_{j,N-j} [ G_j(u,t) G_{N-j}(u,t)- \nonumber \\&-& 
G_N(u,t) ]+{\cal I}_N [ 1 + u - G_N(u,t) ] \nonumber \\
\end{eqnarray}
with the initial condition : $G_1(u,t)=1+u$, and the normalization condition :
$G_N(u=0,t)=1$. $G_N(u,t)$ is the generating function of the probability
distribution : $G_N(u,t)=\sum_{n=1}^{\infty} P_N[n;t](1+u)^{n}$.
The partial derivative in (\ref{no1}) is taken at a fixed size $N$.
The sum on the right hand side (rhs) of eq. (\ref{no1})
represents a binary fragmentation of the primary cluster $N$
into the daughter clusters of mass $j$ and $N-j$ respectively.
The second term on the rhs side is the dissipative term which is 
responsible for the inactivation.
One can transform the discrete variable $j$ in (\ref{no1})
into a continuous one $z = j/N$, which varies from 0 to 1.
With this change, the
fragmentation kernel ${\cal F}_{j,N-j}$ is replaced by a splitting function :
${\hat \Phi}_{z,1-z} \equiv N {\cal F}_{j,N-j} $.

In the limit of 'no dissipation', the FIB process yields the rate equations
of gluodynamics in the Next to Next to Leading Logarithm Approximation (NNLLA)
\cite{ref3}. The inactivation term of
FIB model yields a unique prescription of how to obtain
the rate equations of perturbative QCD in the NNLLA, which would contain
the hadronization effects phenomenologically through the inactivation mechanism
of parton cascading.

The time $t$ in eq. (\ref{no1})
arises within the fragmentation and inactivation kernels
, which themselves are
probabilities per unit of $t$. We define
then the time as : $t=T \ln Y$, where $T$ is a constant,
$Y=\ln(N\Theta /Q_0)$~, $Q_0 = $const and $\Theta$ plays
the role of time and orders the sequence of events.
Assuming now that all physical quantities depend only on the
variable $Y$ and not on $N$ and $\Theta$ separately, we
transform (\ref{no1}) into :
\begin{eqnarray}
\label{no4bis}
\frac{\partial G}{\partial Y}(Y,u)&=&\int_{0}^{1} 
{\gamma_0}^{2}(Y){\hat \Phi}_{z,1-z}[G(Y+\log z,u) \nonumber \\&\times&
G(Y+\log (1-z),u)-G(Y,u)]dz \nonumber \\&+&{\cal R}(Y,u) ~ \ , 
\end{eqnarray}
with
\begin{equation}
\label{norbis1}
{\cal R}(Y,u) = {\gamma_0}^{2}(Y) {\cal{I}}(Y) [1 + u - G(Y,u)] ~ \ ,
\end{equation}
where :
${\gamma_0}^{2}(Y) = 2{\pi}^{-1}N_C \alpha_s(Y)=T/Y$,
and $\alpha_s(Y)$ is the QCD running coupling constant.
${\cal{I}}(Y)$ in (\ref{norbis1}), is the inactivation function
written in the new variables. It should be stressed that no cutoff is needed
and the parton cascades are inactivated gradually by the inactivation function 
${\cal{I}}(Y)$. The generalization of 
eqs. (\ref{no4bis}) to include both quarks and gluons can be found in
\cite{zphys}.

\subsection{Fragmentation scenario for $e^+e^-$ collisions?}

Eqs. (\ref{no4bis}) of the dissipative gluodynamics have been applied for the
description of the hadron multiplicity data in $e^+e^-$ collisions \cite{zphys}. 
The available data have been well described assuming the gaussian inactivation
mechanism of the parton cascades. Based on that, it was concluded that the 
multihadron production in $e^+e^-$ collisions resembles the critical
off-equilibrium shattering process and 
the non-perturbative hadronization stage, 
approximated by the inactivation function ${\cal{I}}(Y)$, is slightly 
shifting this process from the critical line into the shattering phase. The
small deviations from the first-scaling (the KNO-scaling in this case) seen in the
data for $\sqrt{s} \leq 100$ GeV is a finite-size effect related to the small
size of an initial jet. The asymptotic region could not be precisely determined in
these studies but, nevertheless,
the limits imposed by the experimental data on the parameters of the inactivation function
provide the upper limit for the borderline of an asymptotic region : 
$\sqrt{s} \sim 10^2$ TeV.

The experimental analysis of the ratio of the average charged
multiplicity to the dispersion in $e^+e^- \rightarrow q\bar{q}$  
events shows that the multiplicity distributions obey the 
first-scaling law \cite{abreu}. This
confirms the relevance of the hadron multiplicity observable in this process.
Unfortunately, the tail of the multiplicity probability 
distribution has not been resolved experimentally, so it is unknown at
present whether $e^+e^-\rightarrow hadrons$ is the critical process at
presently available energies and whether the hadron
multiplicity is the order parameter in this process. Nevertheless, 
even though the experimental data are yet incomplete in many respects,
the fact that such a simple observable as the hadron multiplicity 
is relevant allows to hope that the future
studies at still higher collision energies will solve the enigma of possible
off-equilibrium shattering phase transition \cite{sing11} in the process :
$e^+e^-\rightarrow hadrons$. This specific transition, dominated by the
perturbative regime of the QCD, is extremely interesting even though 
unrelated with the quark-gluon phase transition, which in turn is 
dominated by the strong-coupling regime of the QCD.

\subsection{Aggregation scenario for $p\bar{p}$ and $AA$ collisions?}

Hadron-hadron and nucleus-nucleus collisions show quite distinct
features from those of lepton-lepton collisions. First of all, 
the multiplicity distributions
show a clear deviation from the first-scaling law \cite{ua5}. The $p\bar{p}$
data for multiplicity distributions at $\sqrt{s}=200, 546$ and 900 GeV 
\cite{ua5} can be approximately superposed at around the maximum 
in the variables of the $\Delta$-scaling law with $\Delta=0.9$
\cite{delphiconf}. On the other hand, the strong scaling violations are seen
in the tail for large multiplicities showing that the global scaling of
multiplicity distributions for different $\sqrt{s}$ may not be
possible. In this context, the two-component scenario including soft and
semi-hard classes of events has been discussed \cite{ugo98}. The system
size-dependence of these components cannot be accurately 
determined based on the existing data, so it is at present difficult to
establish a physical meaning of this scenario. 
One should however mention, that
finite systems in the neighborhood of the first-order liquid-gas 
phase transition, show a similar two-component 
behavior in the cluster multiplicity.

Correlation studies have disclosed the deficiency of the Lund Monte
Carlo program PYTHIA containing jets, resonance production and the
Bose-Einstein correlations as final state interactions, 
for the description of the differential density correlation function \cite{brigitte}.
The correlations are both overestimated at high $p_T$ (high multiplicity) and
underestimated at low $p_T$ (low multiplicity) by PYTHIA
\cite{brigitte,albajar}. Significant improvement can be gained assuming
existence of clusters of low $p_T$ \cite{brigitte,gencl,brigitte1}, 
as included in the non-diffractive event generator GENCL \cite{gencl}. 
The picture of hadron-hadron collisions which emerges from 
these phenomenological studies shows the necessity of a combined 
description of hard scattering processes followed by the string fragmentation 
(the perturbative QCD regime), and the formation and decay of 
clusters in the non-perturbative QCD regime. The absence of an adequate
computer program simulating these two aspects of hadron-hadron
(nucleus-nucleus) collisions is presently a principle obstacle, slowing down a
progress in the understanding of the production mechanism of soft hadrons  
in the strong-coupling, long-distance regime of QCD. The theoretical activity
aiming at the development of such program should be accompanied
by experimental studies of the appropriate methods to evidence the existence of
clusters and to study their sizes.

The arguments put forward in favour of the cluster picture refer often to the
SBM \cite{bootstrap}. In this model, the number of species of possible
constituents of the fireballs and the number of species of fireballs 
grow asymptotically like $\sim m^{-\tau}\exp(m/T_0)$, with $\tau = 3$ 
\cite{nahm}. As discussed in chapt. 3, this value of $\tau$ excludes a possible
identification of $T_0$ with the critical temperature of the thermodynamical
quark-gluon phase transition. Another family
of models refers to the spontaneous breakdown of the chiral symmetry of the QCD
vacuum at a critical temperature. In the approximation $m_u \approx m_d
\approx 0, m_s \rightarrow \infty$, the chiral QCD phase transition is of
second order and the order parameter is given in terms of the sigma 
and pion field condensates \cite{wilczek}. In general, we get 
the guidance in designing appropriate models simulating the critical
behavior in finite temperature QCD from the lattice QCD calculations (see
{\it e.g.} \cite{karsch}). In all these models, 
the proper characterization of the soft phase of QCD should involve an
understanding of features of the largest cluster
(fireball) distribution. 

To design the relevant experimental 
observable in $p\bar{p}$ and $AA$ collisions is by far the
most difficult task. At the same time, this objective is the most important
challenge of soft physics which if unsolved will jeopardies the analysis of
different phases of hot hadronic and quark-gluon matter. 
The hadron (baryon) multiplicity distribution may
contain useful information only if the studied system happens to be found at
the critical point, what obviously limits the relevance of this observable.
In this case, the the study of the scale dependence of correlations by means of
the scale dependence of factorial (cumulant) 
moments of the density distribution, could
reveal the underlying fractal structure of the largest cluster \cite{athens}. 
This is for the moment
largely a theoretical possibility because the exact reconstruction of the
geometry of the largest cluster may be even a harder experimental task 
than finding its mass event by event \cite{hwa}.

\section{CONCLUSIONS}

There is still a long way to understand the multihadron-production processes 
and, in particular, those aspects of these processes which are related to the
strong-coupling long-distance regime of QCD. Experimental studies of
correlations in small domains of the phase-space, posed the problem of the
information contained in the multiparticle fluctuations of produced hadrons
and, more generally, in the fluctuations of observables. We have shown that
these fluctuations have universal features even in small finite systems,
independently of whether the studied process is classical or quantum,
equilibrium or non-equilibrium, dissipative {\it etc.} The information
contained in the particle density and correlations can be extracted by
determining (i) the scaling features (the $\Delta$-scaling law) of the measured
probability distributions of the observable, in particular, the asymptotic
properties of the scaling function, (ii) the anomalous dimension, and (iii) the
scaling features of multiparticle correlations (clustering) in small
phase-space cells. In this way, the relevance of the observable
can be tested and/or the constraints on the multihadron-production
mechanism can be determined.

Much of the experimental search for the scaling features in 
the multiparticle distributions
has been motivated by the prediction of the KNO-scaling as an ultimate symmetry
of the $S$-matrix in ultrarelativistic collisions. This scaling is however
only a particular case of the general $\Delta$-scaling law which has been
introduced only recently. For that reason, neither a 
systematic analysis of the deviation
from the KNO-scaling in terms of the $\Delta$-scaling, nor the asymptotics
of the scaling functions are available at present.

A crucial problem is the determination of the relevant observable for each 
multihadron-production process, which could
exhibit different phases of the studied process through the  
non-trivial fluctuation pattern and its evolution. Whereas the hadron
multiplicity distribution seems to be relevant for $e^+e^-$ collisions, 
which share many aspects of the fragmentation scenario, this observable 
seems to be of secondary importance in $p\bar{p}$ and $AA$ collisions. 
In these latter collision processes,
it is plausible that the aggregation scenario dominates and, hence, the cluster
measures of the data have to be studied. The development of new
methods of extraction of the 'biggest cluster' event by event, accompanied 
by the development of reliable programs simulating both the perturbative QCD
regime and the clustering aspects in the non-perturbative regime, are
the challenging, urgent problems which will determine the future evolution of
the soft physics and the multiparticle correlations.

\vskip 0.5truecm
\noindent
{\bf Acknowledgements}\\
One of us (M.P.) wish to thank the organizers of TORINO 2000 
for creating a lively and
stimulating atmosphere during the meeting. We wish to thank also 
B. Buschbeck for 
stimulating discussions and for sending us her unpublished results.

\end{document}